\documentclass[a4paper,11pt]{article}
\pdfoutput=1 

\usepackage{jinstpub} 

\usepackage{array}
\newcolumntype{P}[1]{>{\centering\arraybackslash}p{#1}}

\usepackage{graphicx,subfig,caption}

\usepackage{hyperref}

\hypersetup{colorlinks}
 
\usepackage[all]{hypcap}  

\hypersetup{colorlinks,
linkcolor=blue,
filecolor=blue,
urlcolor=blue,
citecolor=blue }

\usepackage{siunitx}

\usepackage{amssymb}

\usepackage{amsthm}

\usepackage{amsmath}  
\usepackage{physics}
\usepackage{booktabs}
\usepackage{lineno}

\title{\boldmath Imaging $^{55}$Fe Electron Tracks in a GEM-based TPC Using a CCD Readout}

\author[1,2]{N. S. Phan,\note{Corresponding author}\note{Present address: Los Alamos National Laboratory, Los Alamos, New Mexico 87545, U.S.A.}}
\author[3]{E. R. Lee,\note{Deceased}}
\author{D. Loomba}

\affiliation{Department of Physics and Astronomy, University of New Mexico,\\
	 210 Yale Blvd NE, Albuquerque, NM 87106, USA}

\emailAdd{nphan@unm.edu}

\abstract{Images of resolved 5.9 keV electron tracks produced from $^{55}$Fe X-ray interactions are presented for the first time using an optical readout time projection chamber (TPC). The corresponding energy spectra are also shown, with the FWHM energy resolution in the 30-40\% range depending on gas pressure and gain.  These tracks were produced in low pressure carbon tetrafluoride (CF$_4$) gas, and imaged with a fast lens and low noise CCD camera system using the secondary scintillation produced in GEM/THGEM amplification devices.  The GEM/THGEMs provided effective gas gains of $\gtrsim 2 \times 10^5$ in CF$_4$ at low pressures in the 25-100 Torr range.  The ability to resolve such low energy particle tracks has important applications in dark matter and other rare event searches, as well as in X-ray polarimetry.  A practical application of the optical signal from $^{55}$Fe is that it provides a tool for mapping the detector gain spatial uniformity.}


\begin{document}
\maketitle
\flushbottom


\section{Introduction}
\label{sec:intro}

We show the first results from a high resolution, high signal-to-noise gas TPC used to optically detect and resolve 5.9 keV electron tracks.
All measurements were taken at pressures in the range of 25-100 Torr, where low energy tracks are long enough to be resolved.  The electron tracks produced by the photoabsorption of 5.9 keV $^{55}$Fe X-rays were optically imaged and resolved using a low-noise, high quantum efficiency CCD (charge-coupled device) camera in conjunction with GEM (gas electron multiplier) amplification.  Previously we showed an $^{55}$Fe spectrum derived from tracks imaged  using a similar detector operating in 100 Torr CF$_4$ \cite{phan}. Here we extend this work to show that optical TPCs, when properly optimized, can resolve individual electron tracks at similar energies.  
This capability enables the reconstruction of various track properties, such as the energy-loss profile ($dE/dx$), which can lead to knowledge of particle identification and directionality.  Although this has been demonstrated for energies $\le$$6$ keV using charge readouts (e.g., using pixels or strips) \cite{Bellazzini, Costa, black1, black2}, the lowest energy tracks resolved using optical readouts have been in the 15-22 keV range \cite{sakurai1, sakurai1b, sakurai2}.

Reconstructing low energy recoil tracks is important for a number of applications, such as directional dark matter \cite{spergel} and X-ray polarization \cite{Bellazzini, Costa} experiments, where information sought about the incident particle is contained in the properties of the particle tracks produced in the interaction. For directional dark matter experiments, the measured energy-loss along the track is used for background discrimination \cite{buckland, drift1}, and to search for the directionality signature imprinted on nuclear recoils resulting from galactic dark matter interactions \cite{drift2, drift3}.  In X-ray polarimetry, the angular distribution of photoelectrons produced in the absorption of X-rays from astrophysical sources is used to determine their polarization \cite{Bellazzini, Costa, black1}.  Here, the angular distribution of the electrons is determined by measuring an asymmetry in the energy-loss along their tracks.  Due to the nature of the sources for both applications, detecting and resolving tracks of electron and nuclear recoils down to the lowest energies, $< $$10$ keV, greatly improves the experimental sensitivity.

The stringent requirements for these applications have led to the low pressure gas TPC technology as the basis for most experimental efforts. The advent of Micro-Patterned Gas Detectors (MPGDs) have enabled high gas gains with fine-grained charge or optical readouts, thereby providing the necessary segmentation to image short tracks.  MPGDs with strip \cite{black2} or pixel \cite{Bellazzini, Costa, black1} charge readouts have been used to successfully resolve electron tracks in 2D or 3D at or below $^{55}$Fe energies.  With optical readouts, a variety of methods have been applied to detect and image recoil tracks using the scintillation light produced in the avalanche region for certain gases.  The simplest of these use a photomultiplier tube, which measures the light intensity and, using the pulse shape, one component of the recoil track \cite{Margato1, Margato2}. To image and resolve 2D tracks, optical systems consisting of a CCD camera and some combination of lenses, image intensifiers and fiber optic plates have been used; e.g. \cite{buckland, AustinRamsey,  fraga3, DMTPC0, phan}.  For example, with a capillary gas proportional counter coupled to an image intensified CCD through a lens system  \cite{sakurai2}, 2D electron tracks have been optically resolved  down to $\sim$15 keV. With recent advances, CMOS technology has provided an attractive alternative  to CCDs in certain applications, and its use is also being explored to image and resolve low energy tracks in gaseous TPCs \cite{costa}.

The detector used for our measurements was a TPC filled with high-purity CF$_4$ gas, with GEMs and thick GEMs (THGEMs) used to provide gas amplification and the scintillation light captured by a CCD-based optical readout system. GEMs are a micro-pattern amplification device invented by Sauli at CERN \cite{sauli}; further information on the operation of GEMs with CCD readout can be found in Refs.~\cite{fraga3, fraga4, fraga1, phan}.  THGEMs are similar to GEMs but with dimensions (thickness, hole size, and pitch) that are typically $\geq 3$ times larger.  Exceptionally high gas gains have been achieved in both types of amplification devices, but THGEMs have been shown to excel down at the low pressures of interest here \cite{shalem}.

\begin{figure*}[]
	\centering
	\subfloat[Detector Vessel]{ \includegraphics[width=0.6\textwidth]{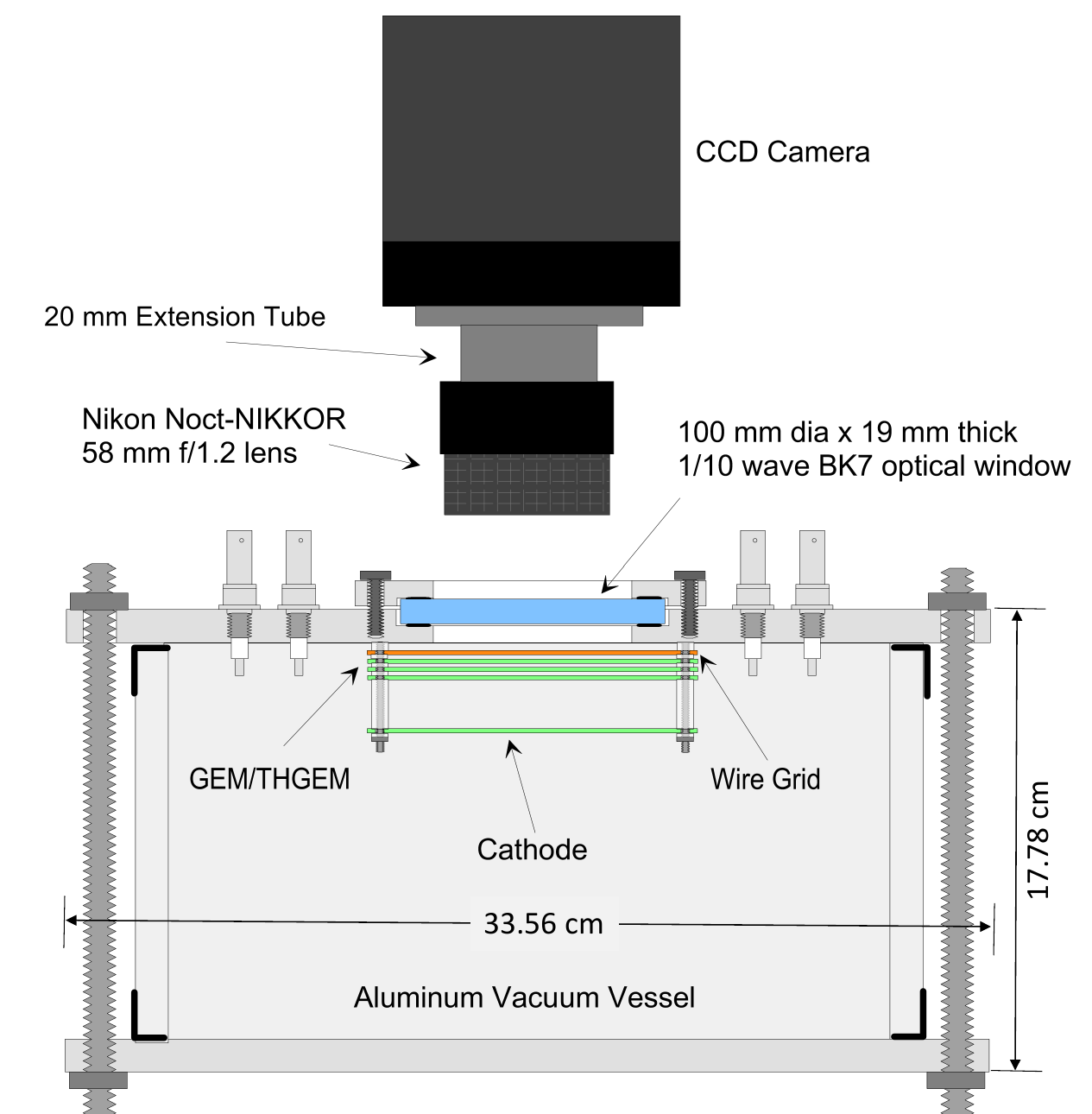}
		\label{fig:detectorvessel}}
	\qquad
	\subfloat[Detection Volume]{\includegraphics[width=0.6\textwidth]{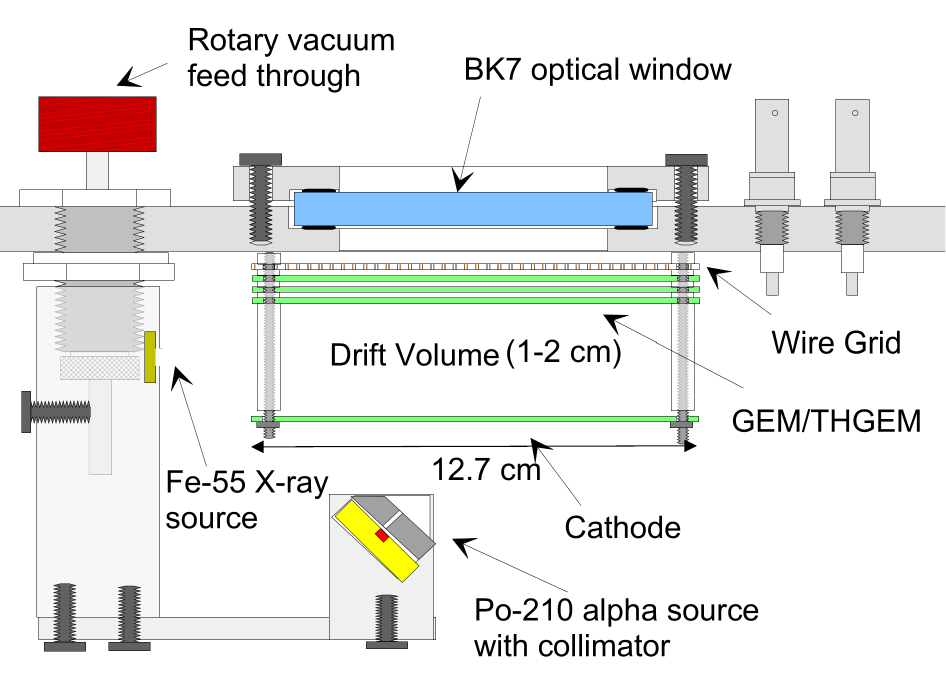}
		\label{fig:detectionvolume}}
	\caption{(a) A drawing of the optical TPC detector showing the aluminum vacuum vessel and CCD camera setup, excluding the rotary feedthrough, camera mount, and calibration sources for clarity.  The light blocking box is also excluded to show the lens and extension tube.  (b) A close up view of the detection volume, showing the locations of the calibration sources, cathode, wire grid, and GEMs/THGEMs. }
	\label{fig:detector}
\end{figure*}

\section{Detector Setup}
\label{sec:setup}

A schematic of the detector setup is shown in Figure~\ref{fig:detector}.  Measurements were made at four pressures, 25 Torr, 35 Torr, 50 Torr and 100 Torr. At each pressure, the choice of amplification device between single GEM/THGEM or multiple GEMs/THGEMs was made to maximize stability, gas gain, and spatial resolution.  For the 100 Torr measurements, the detector used a cascade of three standard copper GEMs separated by 2 mm (Figure~\ref{fig:detector}).  The GEMs were manufactured at CERN using a 50 \si{\micro\meter} thick, $7$ $\times $ 7 cm$^2$ sheet of kapton.  The sheet was copper-clad on both surfaces ($\sim$5 \si{\micro\meter} thick) and chemically etched with a hexagonal array of bi-conical holes of diameter of 50/70 \si{\micro\meter} (inner/outer), at a pitch of 140 \si{\micro\meter}.  For a thorough review of GEMs, see Ref.~\cite{buzulutskov}.  A $7$ $\times $ 7 cm$^2$ cathode was located 1 cm below the GEM stack, defining the detection/drift volume.  The cathode was constructed from a $\sim$ 360 $\mu$m thick copper mesh with  $\sim$500 \si{\micro\meter} pitch.  Finally, a 1 mm pitch anode wire grid plane made from 20 \si{\micro\meter} thick gold plated tungsten wires was situated 3 mm above the top-most GEM (GEM 3), forming the induction gap (further details of this detector can be found in Ref.~\cite{phan}).  The presence of the wire plane creates a well-defined field region above the GEM surface and was empirically found to provide better electrical stability. The geometry of this setup, together with those used for the other pressures is given in Table \ref{tab:config1}.

At pressures below 100 Torr we replaced the triple-GEM cascade with THGEMs, which are better suited for high gain operation at low pressures as noted by Ref.~\cite{shalem}, and confirmed by our experience here.  Even going from 100 Torr to 75 Torr, for example, the maximum stable gas gain in the triple-GEM detector was no longer sufficient to image the electron tracks from $^{55}$Fe interactions.  In addition, due to the lower ionization (scintillation) density of the longer tracks at lower pressures, the gas gain needs to be higher to maintain the same signal-to-noise in the CCD.\footnote{As the gas pressure is lowered a given energy electron track will be longer, covering a larger area of the CCD.  Thus, with the same gas gain, the same number of photons is emitted from the larger area, which results in a reduced light intensity imaged by each pixel of the CCD camera.}
 
All THGEMs used in our studies were fabricated at CERN using a 0.4 mm thick PCB with $\sim$0.3 mm holes mechanically drilled in a hexagonal pattern with a pitch of $\sim$0.5 mm.  To eliminate burring from the drilling process, an annular region of thickness 0.05 mm was chemically etched around each hole. 

In our 50 Torr measurements a single THGEM with an active region of $3 \times 3$ cm$^2$ was used. For 35 and 25 Torr a double-THGEM, with a $9.5 \times 9.5$ cm$^2$ active area\footnote{Two $3 \times 3$ cm$^2$ THGEMs were not available.}, was required to achieve the necessary gas gain. For all three of these lower pressures we also increased the drift gap from 1 to 2 cm to ensure that most of the electron tracks, which are close to 0.5 cm long in 25 Torr (see Figure\ref{fig:fe55-25Torr}), are fully enclosed in the drift volume. The geometrical configurations used at each pressure are given in Table \ref{tab:config1}.  Although we followed the general principles from Ref.~\cite{shalem} for operating THGEMs at low pressures, our geometries were arrived at empirically. Our goal was to achieve the necessary conditions to image and resolve the electron tracks at each pressure. We emphasize that these settings are not unique, and that a dedicated study would likely find better settings that yield higher gas gains.

For all measurements, the detector was housed inside a $\sim$10 liter cylindrical aluminum vacuum vessel.  Prior to powering on the GEMs/THGEMs, the vacuum vessel was pumped out to $\sim$$10^{-3}$ Torr for at least one day before back-filling with high purity (99.999\%) $\text{CF}_{4}$ gas.  A 4-inch diameter BK-7 glass window was positioned above the wire grid to allow scintillation light from the final amplification stage to be transmitted to and imaged by the lens and CCD camera.  BK-7 glass is a low-cost option to quartz that has a relatively high transmittance ($\ge90\%$ in the 350-2100 nm range) for the optical component of the $\text{CF}_{4}$ scintillation, which peaks around 620 nm \cite{morozov, kaboth}.

\begin{table*}[ht]
	\renewcommand{\thetable}{\arabic{table}a} %
	\centering
	\caption{Experimental configurations}\label{tab:config1}
	\setlength{\extrarowheight}{.2em}

\small
 		\begin{tabular}{P{.5cm}P{.8cm}P{2.5cm}P{1.5cm}P{2.2cm}P{2.2cm}P{2.2cm}}
 		\toprule[1.1pt]
		\# & P $\left[Torr\right]$ & Amplification Device & Drift Gap $\left[cm\right]$ & Transfer Gap 1 $\left[cm\right]$ & Transfer Gap 2  $\left[cm\right]$ &  Induction Gap $\left[cm\right]$ \\
		\hline
		\#1  & 100 & 3-GEMs  &  1 & 0.2  &  0.2  &  0.3   \\
		\#2 & 50 & 1-THGEM    &   2 &   -- &     --   &  0.7    \\
		\#3 & 35 & 2-THGEMs  &  2 & 0.4  &  -- &  0.95  \\
		\#4 & 25 & 2-THGEMs  &   2 & 0.65  &  --  &  0.65   \\	
		\bottomrule[1.1pt]

	\end{tabular}
\bigskip

	 \addtocounter{table}{-1}
	 \renewcommand{\thetable}{\arabic{table}b}

	\centering
	\caption{GEM parameters for the configurations shown in Table \ref{tab:config1}}
	\setlength{\extrarowheight}{.2em}

\small
 		\begin{tabular}{P{.2cm}P{.8cm}P{1.5cm}P{.9cm}P{1.45cm}P{1.45cm}P{1.45cm}P{1.1cm}P{1.1cm}P{1.1cm}}
 		\toprule[1.1pt]
		\# & P & Gas Gain & $E_{\mathrm{Drift}}$  & $\Delta V_{\mathrm{(TH)GEM1}}$  &  $\Delta V_{\mathrm{(TH)GEM2}}$ & $\Delta V_{\mathrm{(TH)GEM3}}$ & $E_{12}$ &  $E_{23}$  &  $E_{I}$ \\

 & $\left[Torr\right]$ & &  $\left[V/cm\right]$ & $\left[V\right]$ & $\left[V\right]$ & $\left[V\right]$ & $\left[V/cm\right]$ & $\left[V/cm\right]$ & $\left[V/cm\right]$ \\

		\hline
		\#1a & 100  &  $\sim 1 \times 10^5$ & 400  &  279  &  334   &  380 & 1400 & 1670 & 260 \\
		\#1b & 100  &  $\sim 2 \times 10^5$ & 400  &  290  &  290   &  450 & 1450 & 1450 & 360 \\
		\#2 & 50    &   $\sim 1.5 \times 10^5$ &  200 &    830   &  --  & -- &  --  & -- & 824  \\
		\#3 & 35  &  $\sim 1.6 \times 10^5$ &200  & 573 &  470 &  -- & 718 & -- & 495 \\
		\#4 & 25  &   $\sim 3 \times 10^5$ & 200  & 450  &  650  &  -- & 709 & -- & 682 \\
		
		\bottomrule[1.1pt]
	\end{tabular}
	\label{tab:config2}
\end{table*}

The optical system consisted of a fast 58 mm f/1.2 Nikon Noct-NIKKOR lens coupled to a back-illuminated Finger Lakes Instrumentation (FLI) CCD camera (MicroLine ML4710-1-MB) through a 20 mm extension tube for close-focus imaging.  The whole setup was mounted on top of the vacuum vessel (Figure~\ref{fig:detectorvessel}) in a light tight box.  The camera contained an 18.8 mm diagonal E2V sensor with a $1024 \times 1024$ pixel array (CCD47-10-1-353), consisting of 13 $\times$ 13 $\si{\micro\meter}^2$ pixels.  The mid-band coated CCD sensor had a peak quantum efficiency of 96$\%$ at 560 nm and could be cooled down to a stable operating temperature of $-38^{\circ}$C using the built-in Peltier cooler.  Two readout speeds were available, 700 kHz and 2 MHz, with 16-bit digitization and a maximum $16 \times 16$ on-chip pixel binning.  At the lowest operating temperature and slowest readout mode, the read noise was $\sim$10 e$^-$ rms and the dark current was $\sim$0.03 $e^{-}$/pix/sec with $1 \times 1$ on-chip pixel binning.  At our focusing distance of $\sim$10~cm, the CCD-lens system imaged a $\sim$$3$ $\times$ 3 cm$^2$ physical region of the GEM/THGEM surface.  The known pitch of the holes on this surface was used to calibrate the length-scale of the images.

\section{Detector Calibrations}
\label{sec:calib}

\subsection{GEM/THGEM Gain}
\label{sec:gain}

Calibration of the detector was done using internally mounted $^{55}$Fe (5.9 keV X-rays) and $^{210}$Po (5.3 MeV alphas) sources, which could be individually turned on or off using a rotary feed-through (Figure~\ref{fig:detectionvolume}).  The gas gain was determined using an ORTEC 142IH charge sensitive preamplifier to read out the charge signal from the last GEM/THGEM surface. This required calibrating the preamplifier gain (fC/V) by injecting a known charge and measuring the output voltage.  This was done by connecting an ORTEC 448 research pulse generator to the preamplifier via a test input with a built-in 1 pF capacitor, intended for this purpose.  The 5.9 keV $^{55}$Fe X-ray calibration source was then used to determine the effective gas gain from the output voltage signal of the preamplifier.  The conversion of the X-ray created 172 electron-ion pairs on average, which was calculated from the W-value (the average energy per ionization) of 34.2 eV for CF$_4$ \cite{christophorou}.

For each pressure the maximum stable gain was determined iteratively by raising the GEM/THGEM voltages incrementally and testing for stability at each setting. The latter was done by using the $^{210}$Po source to fire highly ionizing alphas into the detection volume.  If no sparks occurred over several hours, then the voltage setting was deemed stable and the procedure repeated.  The GEM/THGEM voltages, transfer fields and induction fields were arrived at empirically. As stated in Section \ref{sec:setup} where we described the geometrical configurations in Table~\ref{tab:config1}, our goal was to achieve the necessary conditions to image and resolve the electron tracks at each pressure, and not to find the best combination of settings.

At a pressure of 100 Torr, a maximum stable effective gain of $\sim$$1 \times 10^5$ was achieved with the experimental parameters given in Table~\ref{tab:config2}, in the row labeled \#1a. There, the voltage differences across the GEM electrodes are labeled by $\Delta V_{\mathrm{GEM1}}, \Delta V_{\mathrm{GEM2}}$, and $\Delta V_{\mathrm{GEM3}}$, the drift field by $E_{\mathrm{Drift}}$, the transfer field between GEM 1 and 2 by $E_{12}$, and between GEM 2 and 3 by $E_{23}$, and the induction field between GEM 3 and the grid by $E_I$.  By adjusting the GEM voltages (see Table~\ref{tab:config2}, row \#1b) we were able to achieve a higher effective gain of $\sim$$2 \times 10^5$.  This setting, however, was not entirely stable under alpha irradiation, which initiated a spark about once per hour.  Nevertheless, we were able to acquire $^{55}$Fe images and an energy spectrum without any sparks at this setting.

The THGEM voltages and the electric fields in the drift, transfer and induction gaps used for the 50, 35 and 25 Torr measurements are also given in Table~\ref{tab:config2}. For all of these measurements the induction field was reversed so that all electrons produced in the avalanche were collected by the top surface of the final THGEM in the avalanche stage. Contrary to the triple-GEM 100 Torr measurements, where the anode grid voltage was more negative than the final THGEM electrode, this configuration provided better stability at lower pressures with the THGEMs.

For the 50 Torr measurements (\#2) we estimate the gas gain to be $\sim$$1.5 \times 10^5$.  The stability at this voltage against sparking  was similar to that of the highest gain setting at 100 Torr (\#1b), in that $\sim$1 spark/hr was observed when irradiated with alphas.  However, unlike for the high gain 100 Torr case, we found an increase in instabilities after data taking commenced at 50 Torr.  As a consequence we took only a small sample of $^{55}$Fe data (see Section~\ref{sec:lowpressure}).

For the 35 Torr measurements (\#3) the gas gain was $\sim$$1.6 \times 10^5$, which was found to be stable throughout data taking. At 25 Torr (\#4) the gain is estimated to be $ \sim$$3 \times 10^5$, but due to occasional discharges only a small amount of $^{55}$Fe data was taken.

\begin{figure*}[]
	\centering
	\subfloat[$^{55}$Fe tracks in 100 Torr CF$_4$]{ \includegraphics[width=0.42\textwidth]{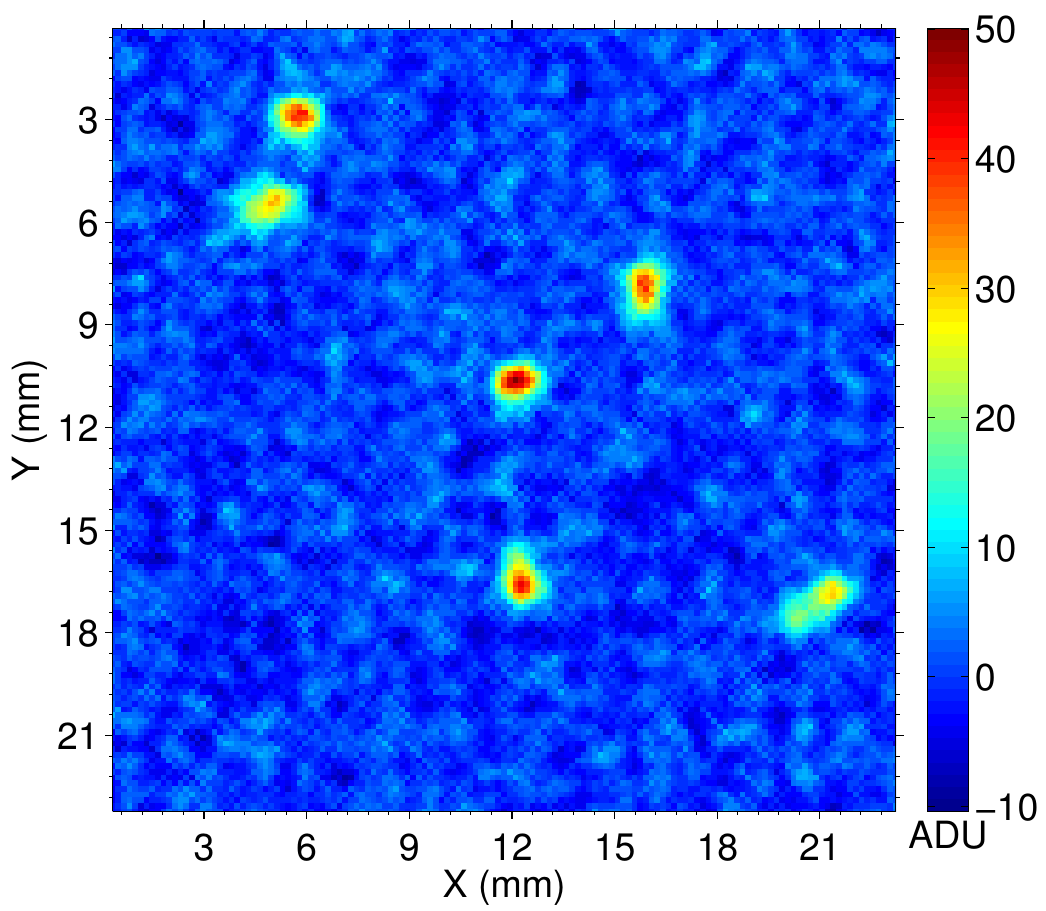}
		\label{fig:fimage100Torr}}
	\qquad
	\subfloat[$^{55}$Fe energy spectrum at 100 Torr CF$_4$]{\includegraphics[width=0.5\textwidth]{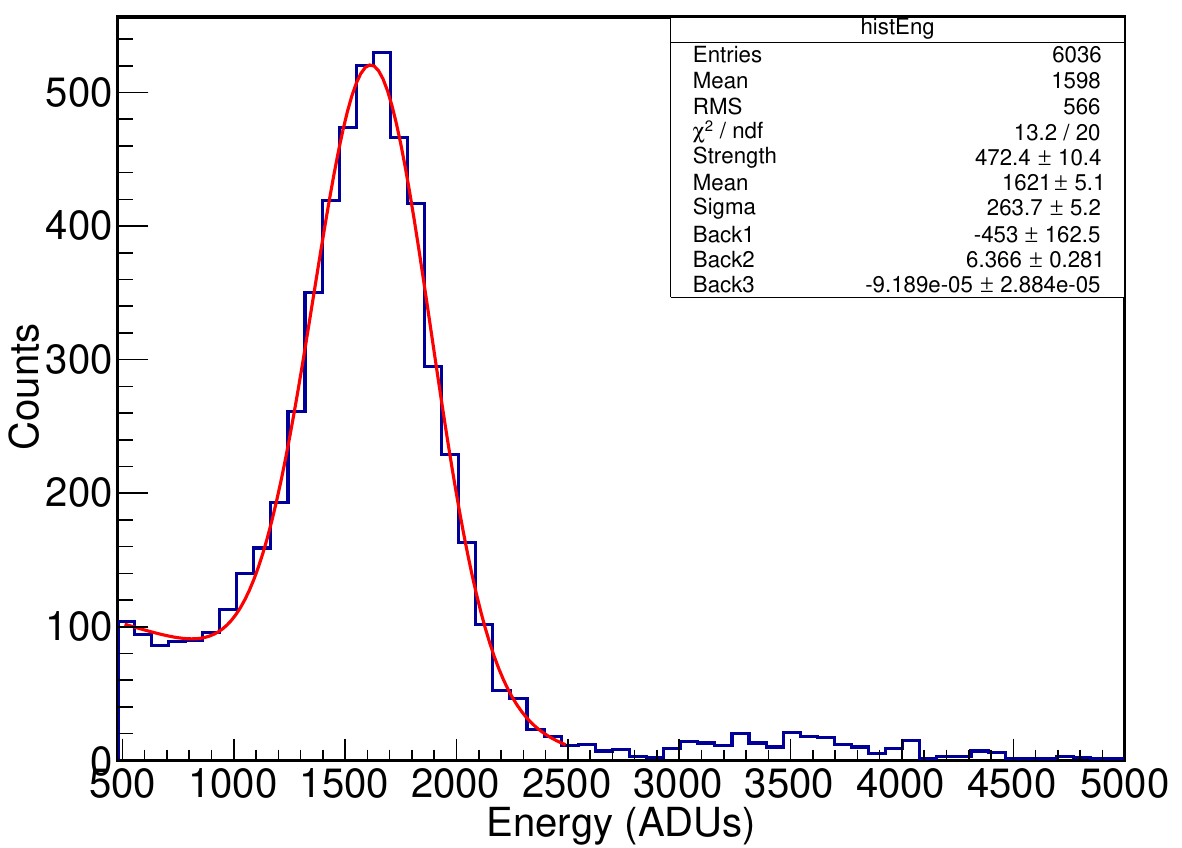}
		\label{fig:fspectrum100Torr}}
	\caption{(a) An image of electron tracks from $^{55}$Fe X-ray interactions (henceforth referred to as $^{55}$Fe tracks) acquired at $ 6\times6$ on-chip binning (pixel scale of 165 \si{\micro\meter}/pix) in 100 Torr CF$_{4}$ with an averaging filter of block size $5 \times 5$ applied to the image to enhance signal-to-noise.  The image is captured at the maximum stable gas gain of $\sim$$10^5$ and has a pixel scale of 165 \si{\micro\meter}/pix.  (b) An energy spectrum of $^{55}$Fe obtained optically from CCD imaging of electronic recoil tracks at $ 6 \times 6$ on-chip binning and the maximum stable gain.  The data is a combination of the start and end data sets in the day eight run (see Section \ref{sec:LY}  and Figure~\ref{fig:peak_vs_days}).  The smaller secondary feature to the right of the primary peak is the result of event pile-up.}
	\label{fig:fe55spectrum100Torr}
\end{figure*}

\begin{figure*}[]
	\centering
	\subfloat[$^{55}$Fe tracks in 100 Torr CF$_4$]{ \includegraphics[width=0.42\textwidth]{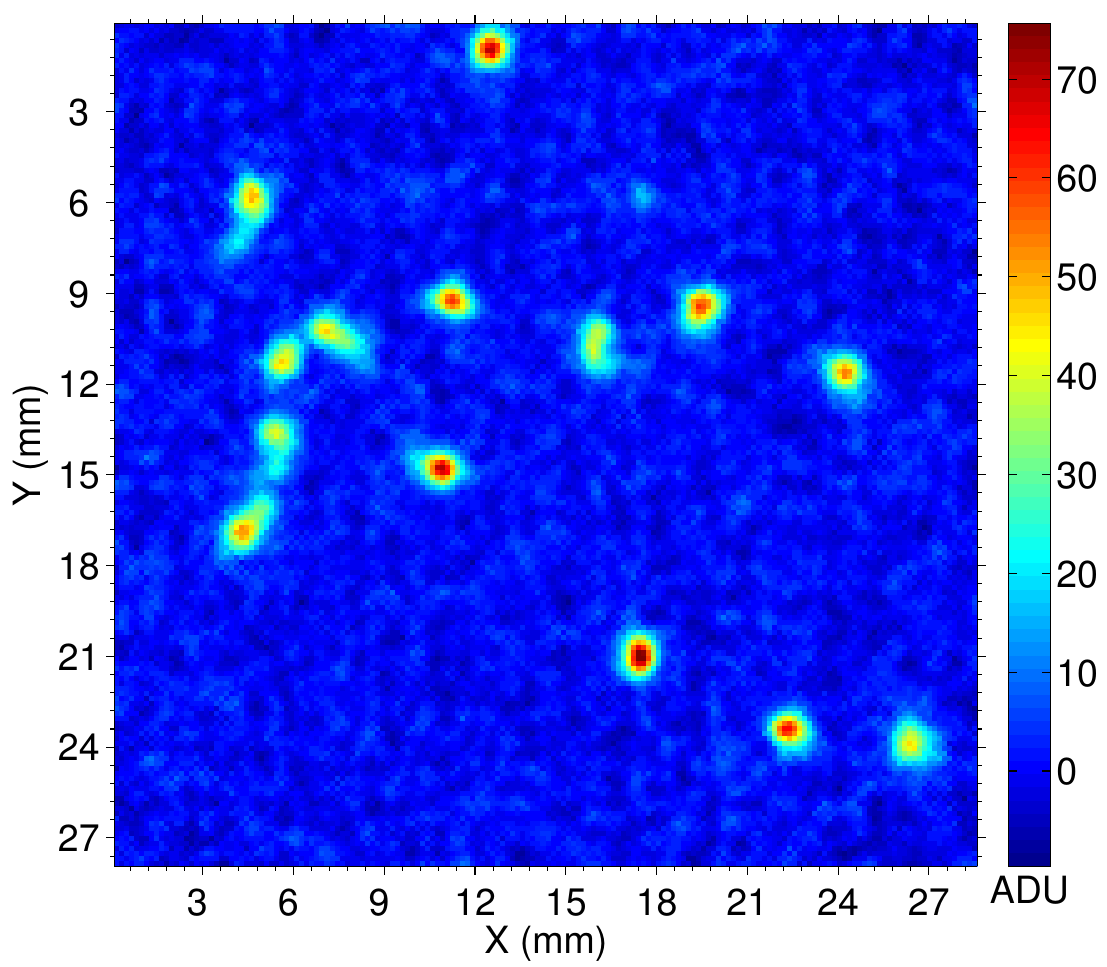}
		\label{fig:fimage100Torrmaxgain6x6}}
	\qquad
	\subfloat[$^{55}$Fe energy spectrum at 100 Torr CF$_4$]{\includegraphics[width=0.50\textwidth]{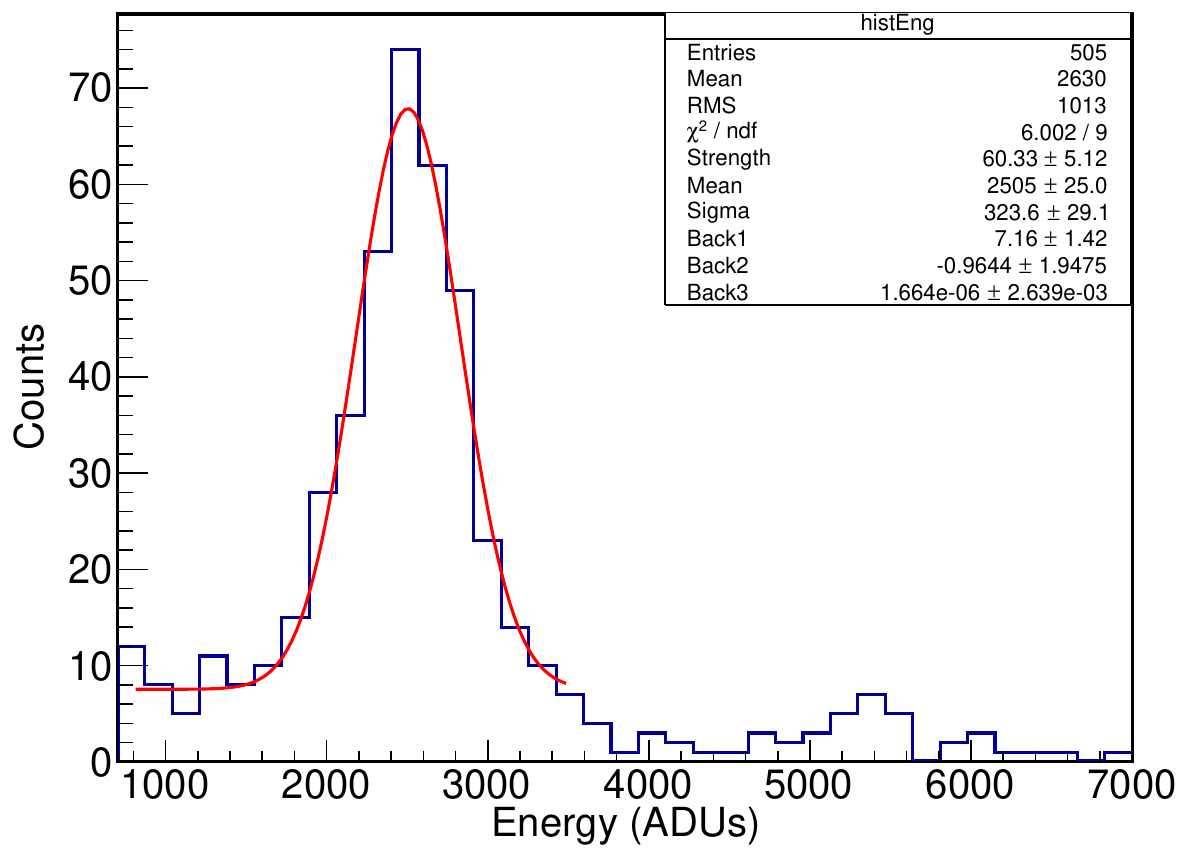}
		\label{fig:fspectrum100Torrmaxgain6x6}}
	\caption{(a) An image of $^{55}$Fe tracks acquired at $6\times 6$ on-chip binning in 100 Torr CF$_{4}$ with an averaging filter of block size $5 \times 5$ applied to the image to enhance signal-to-noise.  The image is captured at the maximum gas gain of $\sim 2 \times 10^5$ and shows that even in 100 Torr, $^{55}$Fe tracks are resolved.  (b) An $^{55}$Fe energy spectrum obtained optically from CCD imaging of electronic recoil tracks at $ 6 \times 6$ on-chip binning and maximum gain of $\sim 2 \times 10^5$.  The smaller secondary feature to the right of the primary peak is due to event pile-up.}
	\label{fig:fe55100Torrmaxgain6x6}
\end{figure*}

\subsection{CCD Calibration}
\label{sec:ccd}

The CCD images (or frames) were calibrated through a standard approach~\cite{Janesick} by using a set of co-averaged flat-field and dark frames.  An exposure time of 5 seconds was chosen for the acquisition of both calibration and data images.  Dark frames, taken with the camera shutter closed, were used to correct for the variable accumulation rate of dark current across the pixels in the CCD sensor.  Flat-field frames, used to correct for vignetting and pixel to pixel variation in light sensitivity, were acquired by taking exposures of a uniformly illuminated screen.  For each type of calibration frame, a set was acquired and co-averaged together to create a master calibration frame.  For accuracy, the averaging required rejecting pixels in individual frames that suffered direct hits from cosmic rays or radioactivity. This was done using an algorithm that compared the value of the same pixel across the set of frames, excluding those above three sigmas of the initial average of the pixels.  The average was re-computed and the process iterated until there was a convergence in the average value of the pixels.  Finally, the data image was calibrated by subtracting the master dark frame and dividing the resulting frame by the normalized, master flat-field frame.

\section{Results}
\label{sec:Results}

In this Section we derive results from CCD images of $^{55}$Fe tracks taken at each pressure using the experimental configurations given in Tables~\ref{tab:config1} and \ref{tab:config2}.  The CCD images were acquired using a level of on-chip binning determined by balancing the lower read noise at higher binning with the desire for finer spatial granularity imaging at lower binning that is needed to resolve the short electron tracks. We struck this balance by setting the binned pixel-scale\footnote{Pixel-scale is defined as the real space imaged per pixel. For $6 \times 6$ binning this is 165 \si{\micro\meter} per binned pixel.} to match the (TH)GEM hole pitch, which sets a lower limit on how finely a track in our detector can be imaged.  Thus, for the 100 Torr measurements made using the 3-GEM configuration (\#1 in Tables~\ref{tab:config1} and \ref{tab:config2}), we chose $6 \times 6$ binning  because the resulting pixel-scale of 165 \si{\micro\meter}/px  closely matches the GEM hole pitch of 140 \si{\micro\meter}. For the measurements made using THGEMs, a  $16 \times 16$ binning was chosen because its 440 \si{\micro\meter}/px pixel-scale closely matches the $\sim$0.5 mm hole pitch.  Using a large sample of individual $^{55}$Fe tracks obtained at each pressure, the average track-length is measured, and in the case of 100 and 35 Torr, the corresponding energy spectra are also derived. Quantitative results from these measurements are summarized in Table~\ref{tab:summary}. 

To obtain the spectra shown in this section, the images are analyzed using an algorithm developed with MATLAB and its image processing toolbox. First, the calibrated images are binned $4 \times 4$ in software, and pixels above a set threshold of 3.2$\sigma_{im}$ are labeled as objects. Objects containing a pixel(s) on the boundary of the image or have less than four contiguous pixels are rejected.  Once the objects are identified, the binned image is up-sampled back to its original size.  A position and intensity weighted ellipse is fitted for each identified object in the image.  Important properties of the object such as its energy, track length, and width are determined from this fit.  For a more detailed description of the algorithm, refer to Ref.~\cite{phan}.

\subsection{100 Torr}
\label{sec:100Torr}

A sample image containing $^{55}$Fe tracks taken at the maximum stable gain settings in 100 Torr CF$_4$ is shown in Figure~\ref{fig:fimage100Torr}.  On-chip binning of $6 \times 6$ was used for this image, as discussed above, and it can be seen that the signal is well above the noise in the CCD image and individual tracks are resolved.  With this capability, an $^{55}$Fe spectrum can be measured in sub-regions of the GEM and used to characterize the spatial uniformity of gas gain across its surface. This provides an attractive alternative to the use of high flux X-ray generators commonly used for quality control of GEMs (e.g. Ref. \cite{fraga5})

The energy spectrum of the scintillation light from individual $^{55}$Fe electron tracks is shown in Figure~\ref{fig:fspectrum100Torr} in units of ADUs\footnote{Analog to Digital Units, with 1 ADU equal to $\sim$1.3 $e^{-}$ produced in our CCD sensor.}.  The peak value in the spectrum is obtained from a fit using a single Gaussian for the signal component, and a constant plus exponential for the background component.  The parameters labeled Strength, Mean, and Sigma in the information window of the plot correspond to the Gaussian component, whereas the parameter Back1 belongs to the constant component and the parameters Back2 and Back3 belong to the exponential component.  As noted above, the background is the sum of the constant and exponential components.  The range of the fit is set so as to exclude the secondary peak seen at $\sim$3400 ADU, which is due to pile-up events.  The value of the secondary peak is twice that of the primary peak, and this is the value that would be expected when two tracks are overlapping one another.  The fit has a reduced $\chi^2$ $(\chi^2/ndf) = 0.66$, a peak value of $\mu = 1621 \pm 5$, and $\sigma = 264 \pm 5$.  The FWHM energy resolution is 38\% and, setting the peak value from the fit equal to 5.9 keV, we obtain an energy conversion factor of 275 ADUs/keV. The key parameters from the $^{55}$Fe fit are summarized in the first row of Table~\ref{tab:summary}, which is labeled by the experimental configuration \#1 used for the measurements, as described in Tables~\ref{tab:config1} and \ref{tab:config2}. In the last column of Table~\ref{tab:summary} the average track length of 1.9 mm is also listed.

In Figures~\ref{fig:fimage100Torrmaxgain6x6} and \ref{fig:fspectrum100Torrmaxgain6x6}, a sample image of $^{55}$Fe tracks in 100 Torr CF$_4$ at a higher gain ($\sim$2$\times10^5$), also imaged at $6\times 6$ binning, is shown along with the corresponding energy spectrum.  The spectrum is fit using the procedure described above, and the fit parameters summarized in Table~\ref{tab:summary} (configuration \#1b). As expected, the higher  gas gain results in a higher peak ADU value, however, what was unexpected was a significant improvement of the FWHM energy resolution than that of the moderate gain data.

\begin{figure*}[]
	\centering
	\subfloat[$^{55}$Fe tracks in 100 Torr CF$_4$]{ \includegraphics[width=0.42\textwidth]{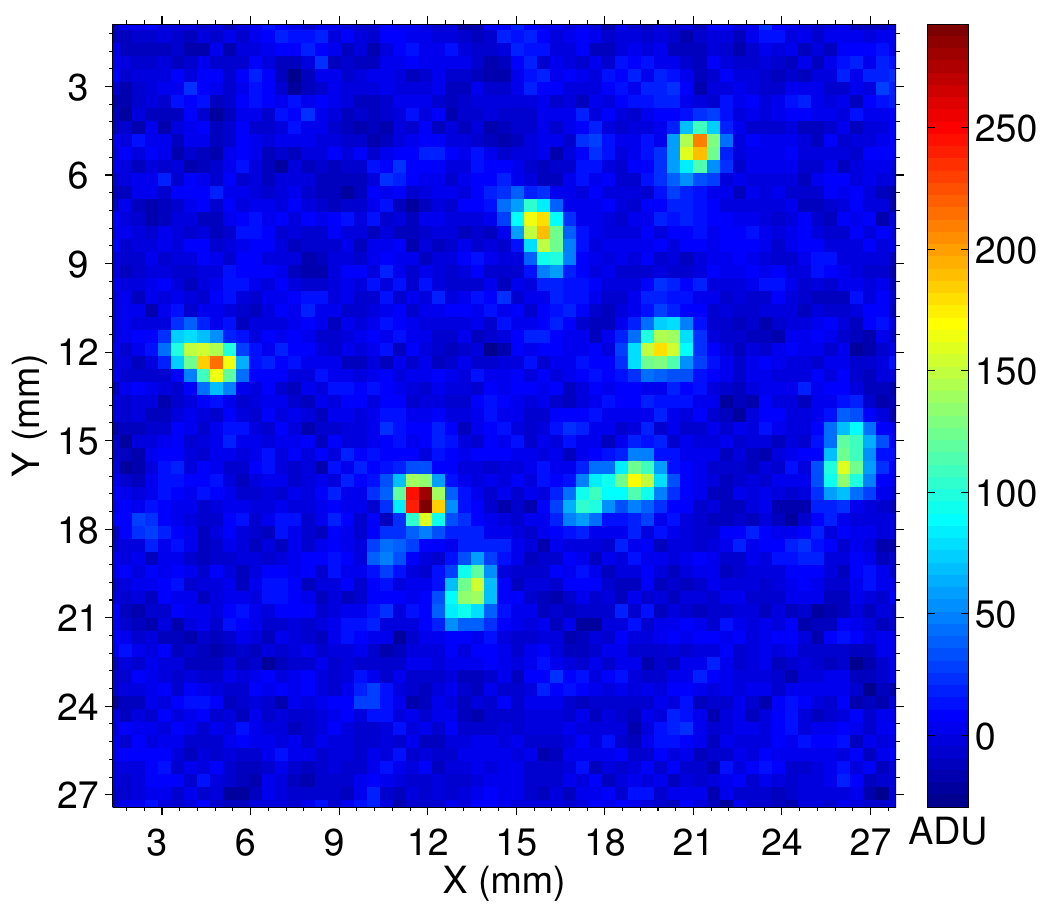}
		\label{fig:fimage100Torrmaxgain16x16}}
	\qquad
	\subfloat[$^{55}$Fe energy spectrum at 100 Torr CF$_4$]{\includegraphics[width=0.50\textwidth]{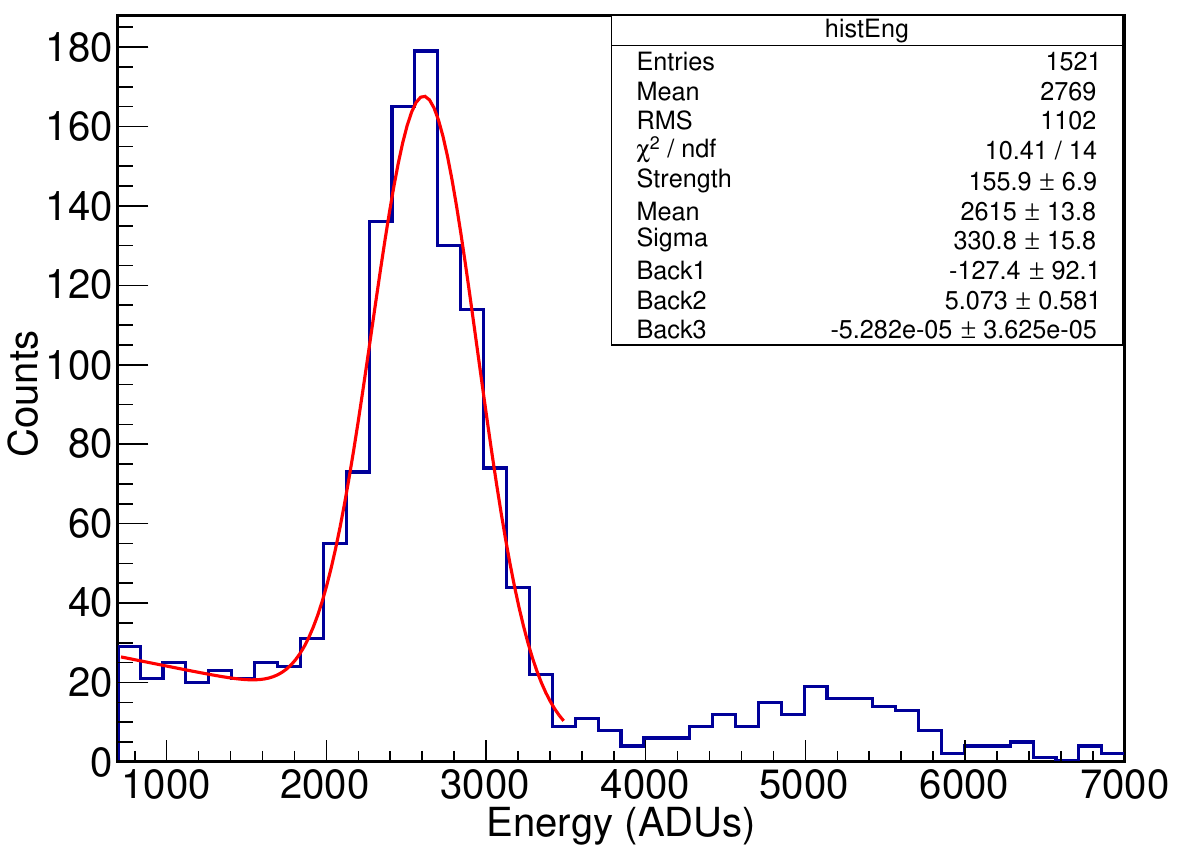}
		\label{fig:fspectrum100Torrmaxgain16x16}}
	\caption{(a) An image of  $^{55}$Fe tracks  acquired at $16\times 16$ on-chip binning (pixel scale of 440 \si{\micro\meter}/pix) in 100 Torr CF$_{4}$ with an averaging filter of block size $3 \times 3$ applied to the image to enhance signal-to-noise.  The image is captured at the maximum gas gain of $\sim 2 \times 10^5$ and shows that even in 100 Torr, low energy 5.9 keV electron tracks are resolved.  (b) An $^{55}$Fe energy spectrum obtained optically from CCD imaging of electronic recoil tracks at $ 16 \times 16$ on-chip binning and maximum gain of $\sim 2 \times 10^5$.  The smaller secondary feature to the right of the primary peak is due to event pile-up.}
	\label{fig:fe55100Torrmaxgain16x16}
\end{figure*}

For both gain settings the average electron track-length, $\sim$2 mm, was significantly longer than the image granularity set by the 165 \si{\micro\meter} pixel-scale. Thus, we also acquired data at the high gain setting with $16 \times 16$ binning (440 \si{\micro\meter}/px), where the contribution from read noise, per real-space area, is lower than $6\times 6$ binning.  A sample image of $^{55}$Fe tracks at this binning and the corresponding energy spectrum are shown in Figures~\ref{fig:fimage100Torrmaxgain16x16} and \ref{fig:fspectrum100Torrmaxgain16x16}. A comparison of the spectrum fit parameters (third row of Table~\ref{tab:summary}) with those of the $6\times 6$ binning shows no change in energy resolution, but a 4\% higher peak ADU value. This could be due to the higher signal-to-noise with $16\times16$ binning, resulting in better efficiency for capturing the entire track, but we cannot rule out the possibility of a drift in gas gain over the time taken between acquiring the two data sets. Although the tracks are still resolved at this coarser binning, its effect is apparent in the average track-length, which is $\sim$25\% longer than for $6\times 6$ binning.

\subsection{25, 35 \& 50 Torr}
\label{sec:lowpressure}

\begin{figure*}[]
	\centering
	\subfloat[$^{55}$Fe tracks in 50 Torr CF$_4$]{ \includegraphics[width=0.42\textwidth]{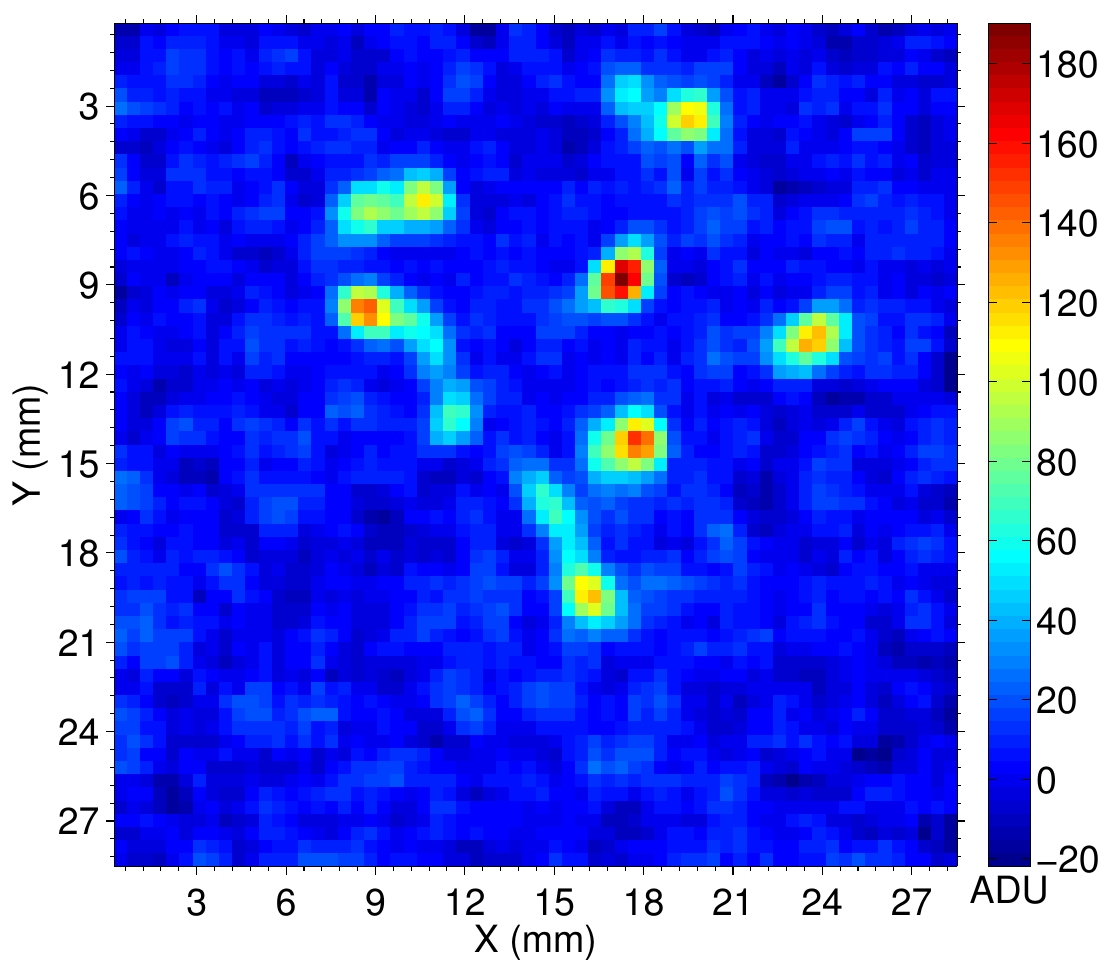}
		\label{fig:f1image50Torr}}
	\qquad
	\subfloat[$^{55}$Fe tracks in 50 Torr CF$_4$]{\includegraphics[width=0.42\textwidth]{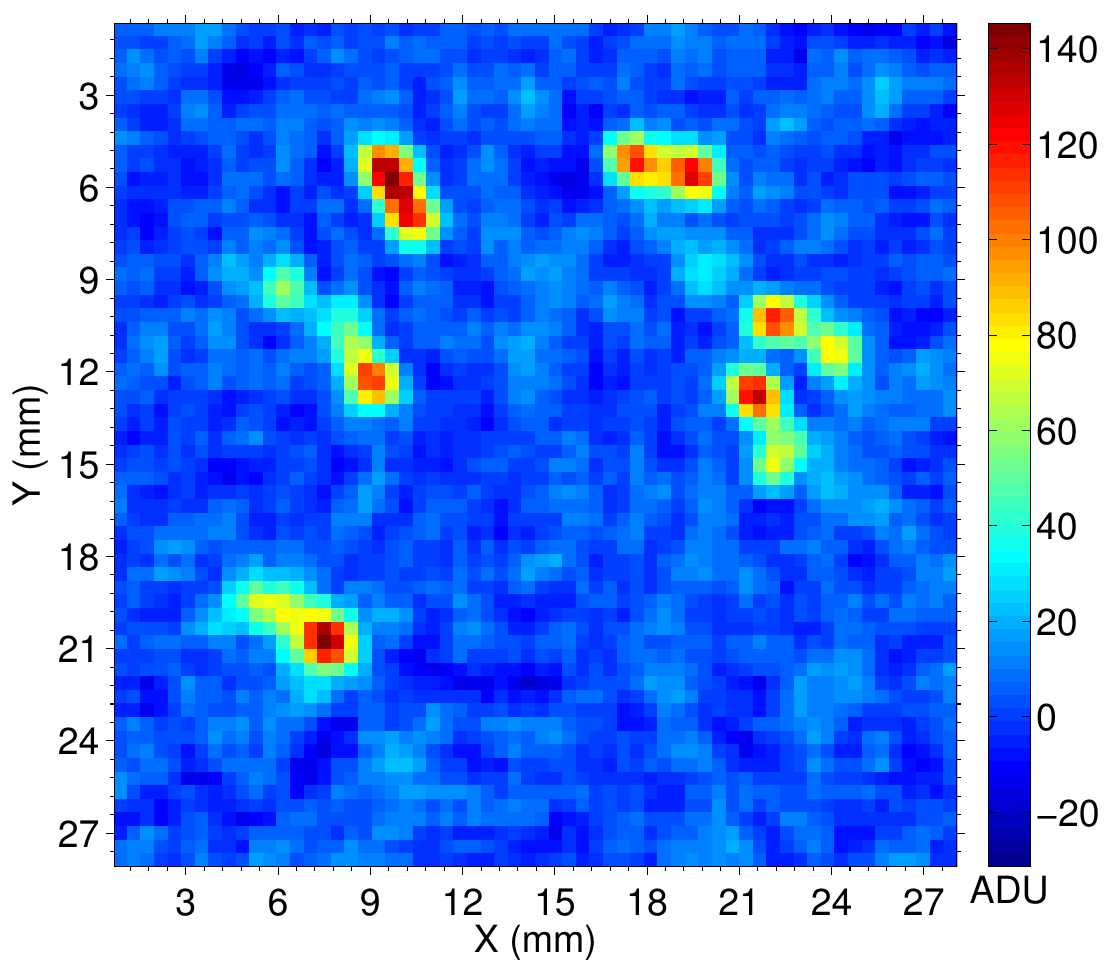}
		\label{fig:f2image50Torr}}
	\caption{(a)-(b) Images of  $^{55}$Fe  tracks in 50 Torr CF$_4$ at $16 \times 16$ on-chip binning.  An averaging filter with a $3 \times 3$ block size has been applied to the image to improve signal-to-noise without significantly degrading resolution.  At this pressure, the tracks are well resolved and fluctuations in energy loss and range straggling are also clearly observable. }
	\label{fig:fe55-50Torr}
\end{figure*}

\begin{figure*}[]
	\centering
	\subfloat[$^{55}$Fe tracks in 35 Torr CF$_4$]{ \includegraphics[width=0.42\textwidth]{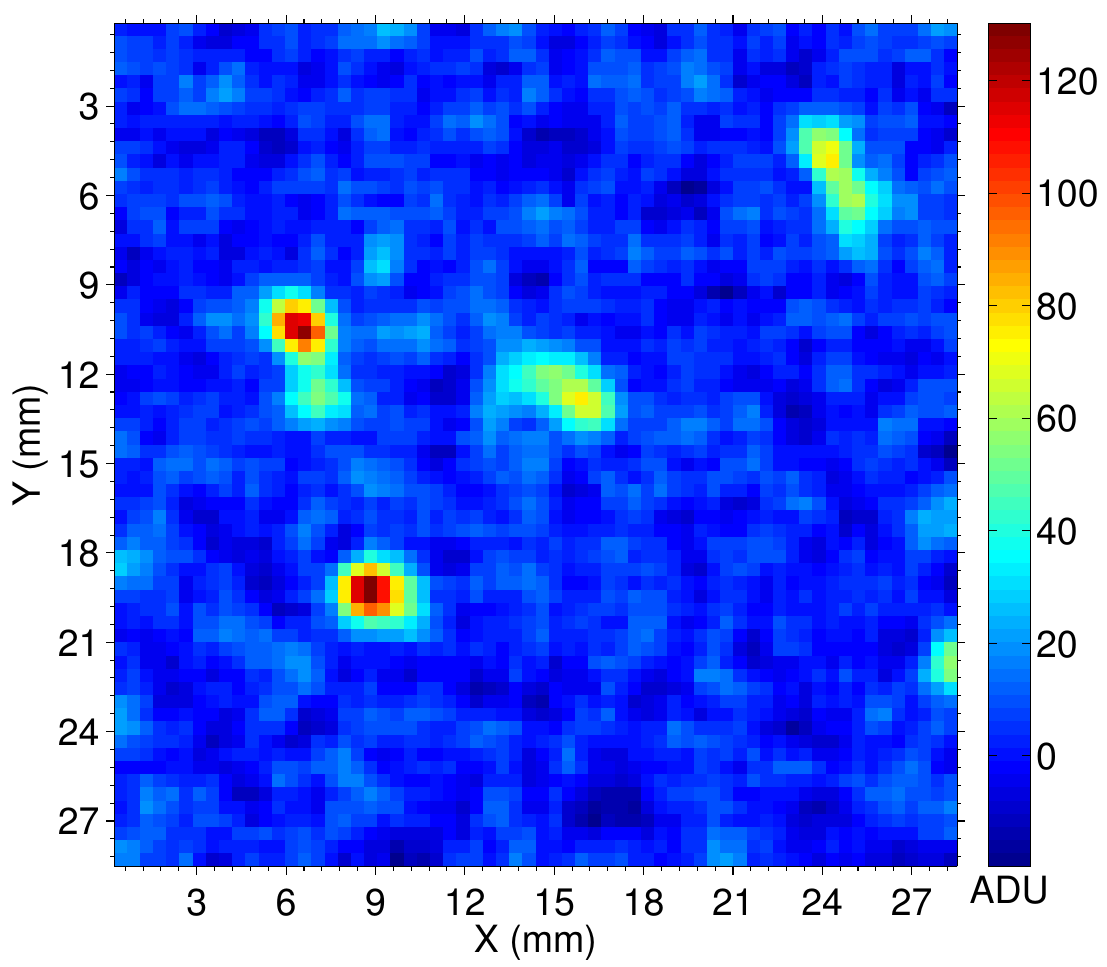} %
		\label{fig:fimage35Torr}}
	\qquad
	\subfloat[$^{55}$Fe energy spectrum at 35 Torr CF$_4$]{\includegraphics[width=0.50\textwidth]{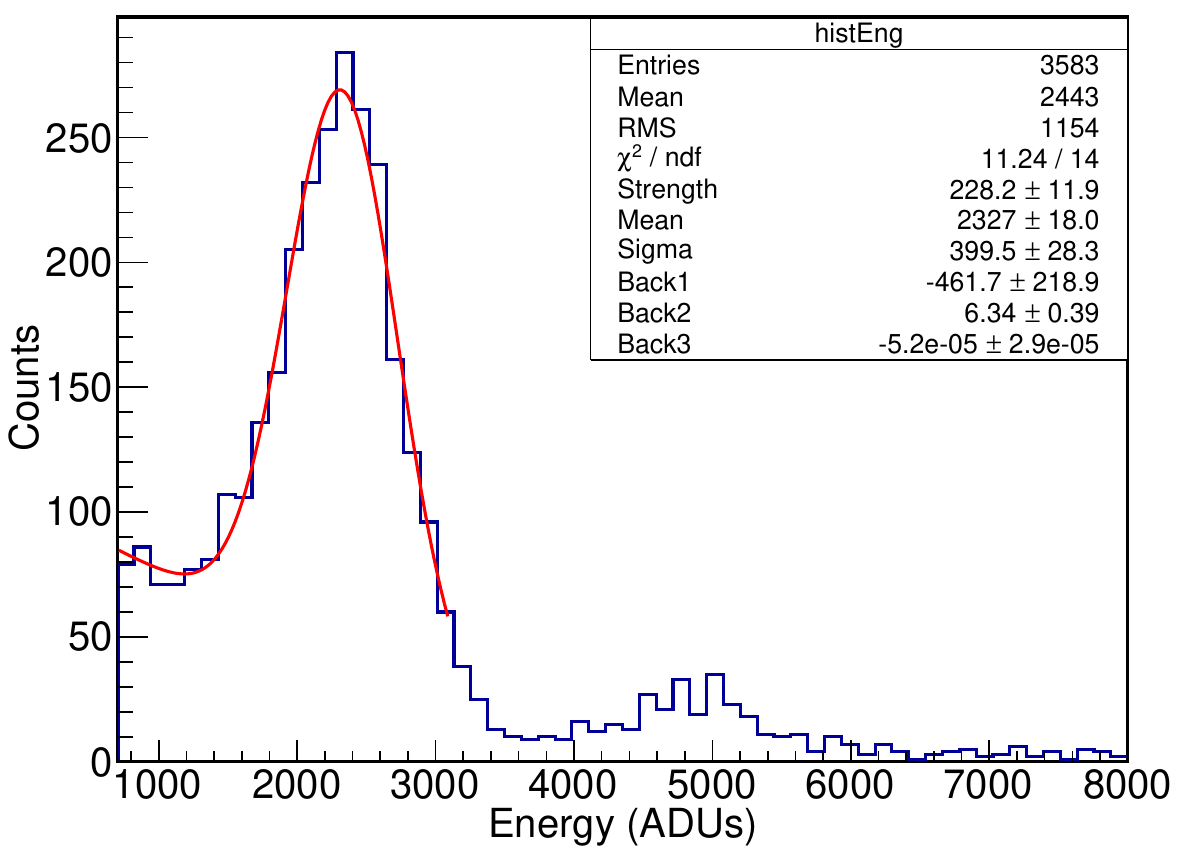} %
		\label{fig:fspectrum35Torr}}
	\caption{(a) An image of $^{55}$Fe tracks in 35 Torr CF$_4$ with an averaging filter applied to enhance signal to noise.  The tracks are clearly resolvable as extended objects rather than diffused points at this pressure.  (b) An energy spectrum obtained from CCD imaged $^{55}$Fe electronic recoil tracks in 35 Torr at $16\times 16$ on-chip binning and maximum stable gain.  The smaller secondary feature on the right of the primary peak is due to event pile-up.}
	\label{fig:fe55spectrum35Torr}
\end{figure*}

In Figure~\ref{fig:fe55-50Torr}, two sample images of $^{55}$Fe tracks taken using a single THGEM at a pressure of 50 Torr in CF$_4$ are shown.  As expected, the tracks are longer ($\sim$3.7 mm on average) and much better resolved than those in the 100 Torr data and, in addition, ionization density gradients along the electron tracks are also clearly visible. In a number of tracks a clear assymmetry in ionization density is seen, which allows one to reconstruct the start (low density) and end (high density) of the electron track.  The corresponding energy spectrum is not shown due to an increase in electrical instabilities, as stated in Section~\ref{sec:gain}.  Instead of risking damage to the THGEM we took just a small amount of data in this configuration.  Nevertheless, as the images in Figure~\ref{fig:fe55-50Torr} unambiguously show, TPCs with CCD-based optical readouts and sufficiently high signal-to-noise and imaging resolution, can {\it{resolve}} electron tracks with energies as low as 5.9 keV.  

For the 35 Torr CF$_4$ measurements 2 THGEMs were required to obtain sufficient gas gain (see Section~\ref{sec:setup} and Tables~\ref{tab:config1} and \ref{tab:config2}).  Figure~\ref{fig:fimage35Torr} shows a sample image containing $^{55}$Fe tracks taken in this configuration.  The energy spectrum obtained from a series of these images is shown in Figure~\ref{fig:fspectrum35Torr}.  A fit of the spectrum used the same functions for signal and background as for the 100 Torr data (Section~\ref{sec:100Torr}) and the key parameters summarized in Table~\ref{tab:summary}. The resulting FWHM energy resolution of 40\% is similar to, within errors, to that obtained from the moderate gain 100 Torr data.

As in the 50 Torr data the tracks in 35 Torr are clearly resolved but the resolution is poorer due to the presence of the second THGEM and the transfer region within the double THGEM amplification structure.  This is expected and is due to two effects. The first is that the holes of the 2 THGEMs are not aligned, so that the charge cloud forming the track will suffer broadening as it passes from THGEM 1 to THGEM 2. For the larger hole pitch, $\sim$0.4 mm, of THGEMs (relative to thin GEMs) this could be a significant effect.  The second effect is due to the additional diffusion of the charge cloud as it traverses the 4 mm transfer gap.  This diffusion is worse, per unit drift length, than in the drift region because the transfer $E$ field is so high. Using MAGBOLTZ \cite{biagi} we estimate that the transverse diffusion in the 0.4 cm transfer gap is $\sigma = 545 ~\mu $m, compared to $\sigma = 465 ~\mu $m in the 2 cm drift region.

\begin{figure*}[]
	\centering
	\subfloat[$^{55}$Fe tracks in 25 Torr CF$_4$]{ \includegraphics[width=0.42\textwidth]{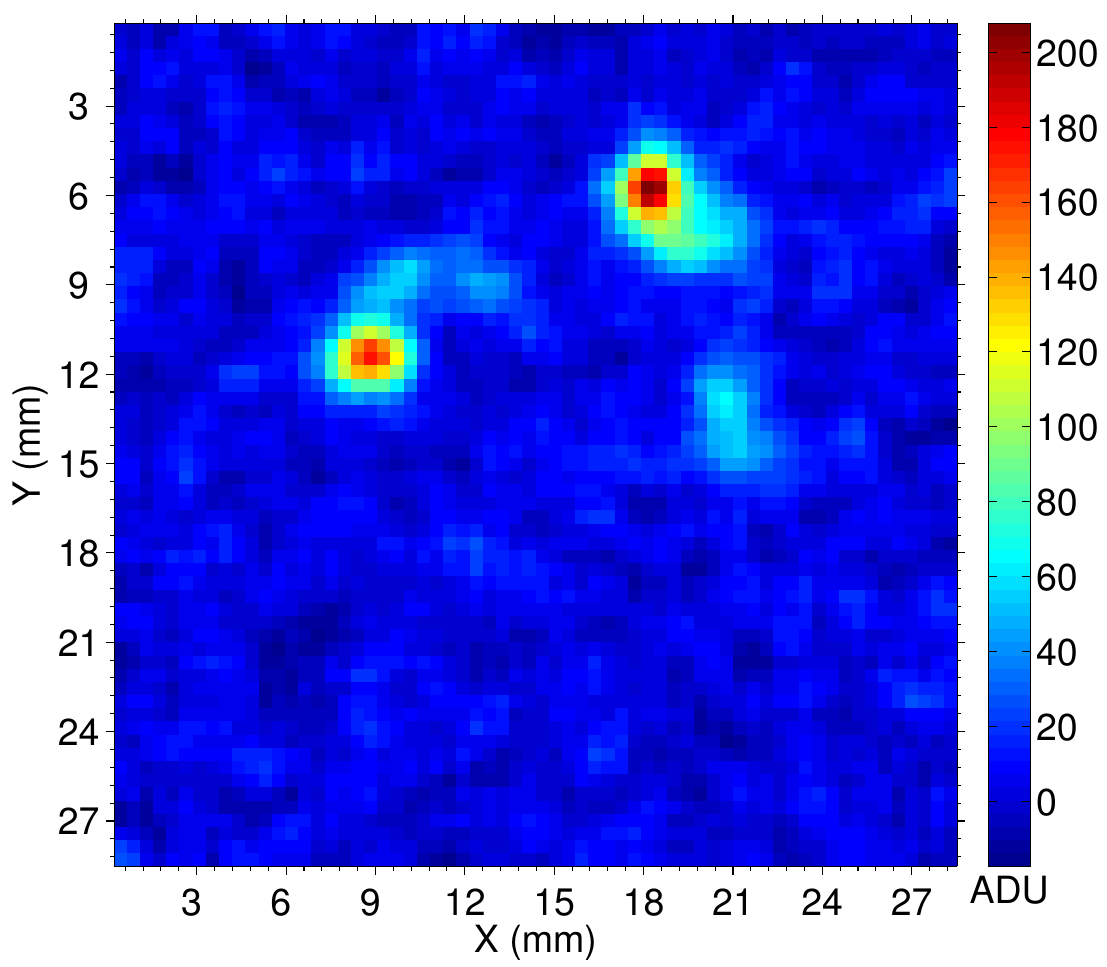} %
		\label{fig:fimage25Torr}}
	\qquad
	\subfloat[$^{55}$Fe tracks in 25 Torr CF$_4$]{\includegraphics[width=0.42\textwidth]{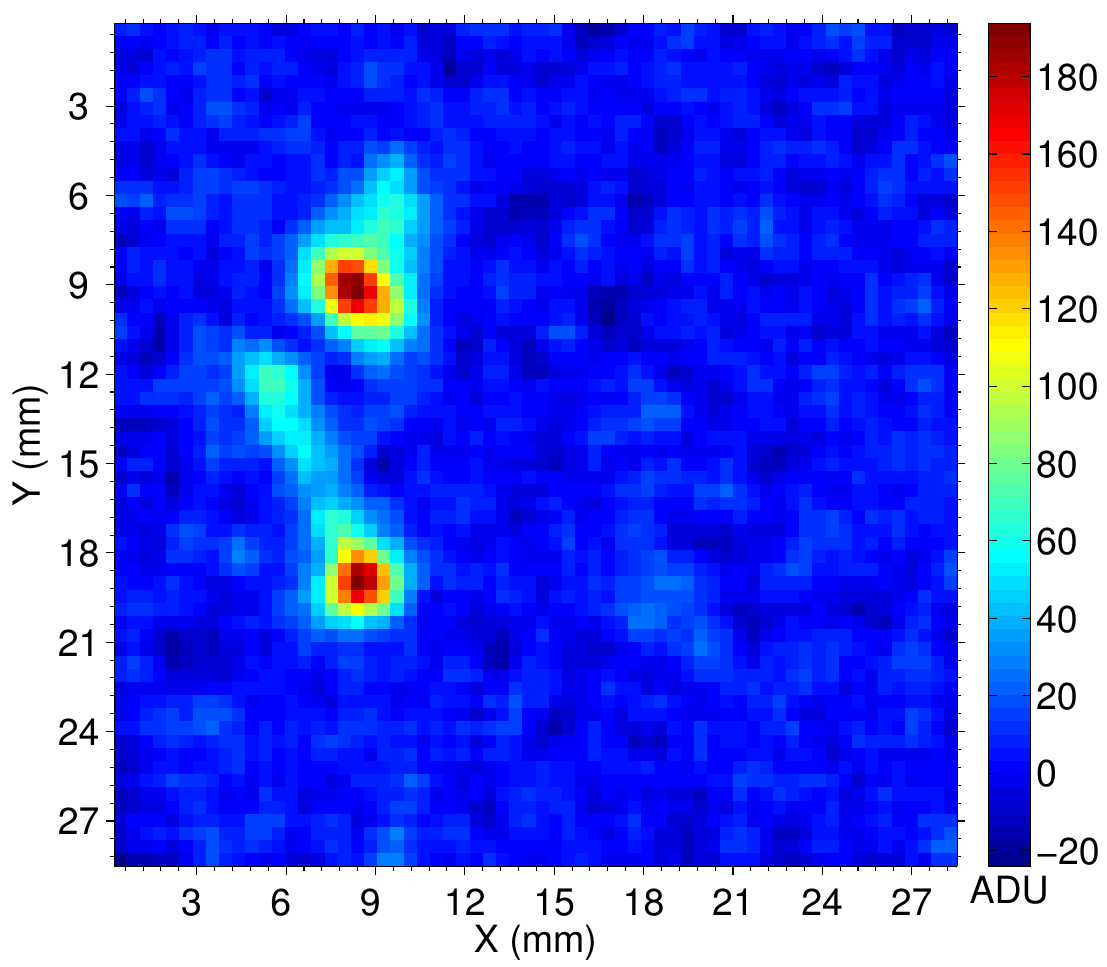} %
		\label{fig:f2image25Torr}}
	\caption{(a)-(b) Sample images of $^{55}$Fe tracks in 25 Torr CF$_4$ using two THGEMs and $16 \times 16$ CCD on-chip binning.  A $5 \times 5$ averaging filter has been applied to enhance signal to noise.  The tracks are resolvable at this pressure and gas gain.}
	\label{fig:fe55-25Torr}
\end{figure*}

Finally, in Figure~\ref{fig:fe55-25Torr} we present sample images of $^{55}$Fe tracks acquired in 25 Torr CF$_4$.  However, there was insufficient data taken for an energy spectrum because electrical instabilities (discharges) prevented a long data run.  But with additional fine tuning of THGEM voltages and the detector setup, it is conceivable that stable operation could be achieved at the necessary gas gains needed to image individual $^{55}$Fe tracks.

\begin{table*}[ht]
	\centering
	\caption{Summary of Results}
	\setlength{\extrarowheight}{.2em}
	\small
 	\begin{tabular}{P{.5cm}P{1.1cm}P{2.cm}P{1.5cm}P{1.6cm}P{1.8cm}P{2cm}P{1.6cm}}

		\toprule[1.1pt]

		\# & P $\left[Torr\right]$ & Amplification Device & Gas Gain & Pixel Binning & FWHM Energy Res. $\left[\%\right]$ & Energy Conv. Factor $\left[ADU/keV \right]$ & Avg. Track Length $\left[mm \right]$ \\

		\hline
		\#1a & 100  & 3-GEMs &  $\sim 1 \times 10^5$ & $6\times6$  &  38  &  275   & 1.9 \\
		\#1b & 100  & 3-GEMs &  $\sim 2 \times 10^5$ & $6\times6$  &  30  &  425   &  2.0 \\
		\#1c & 100  & 3-GEMs &  $\sim 2 \times 10^5$ & $16\times16$  &  30  &  443   &  2.5 \\
		\#2 & 50  & 1-THGEM &  $\sim 1.5 \times 10^5$ &  $16\times16$ &    --   &  --  & $\sim 3.7$ \\
		\#3 & 35  & 2-THGEMs &  $\sim 1.6 \times 10^5$ & $16\times16$  & 40 &  394 &  4.1 \\
		\#4 & 25  & 2-THGEMs &  $> 2 \times 10^5$ & $16\times16$  & --  &  --  &  $\sim 4.6$ \\
		\bottomrule[1.1pt]
	\end{tabular}
	\label{tab:summary}
\end{table*}

A summary of the results for the different data sets and their corresponding detector configurations are found in Table~\ref{tab:summary}.  
Although the average track-length increases with decreasing pressure, the relationship is not linear. We note that this quantity is the detected track length, which depends on multiple factors besides the gas pressure.  These include diffusion, smearing within the amplification stage, gas gain (signal-to-noise), and CCD pixel binning.

\subsection{Light Yield and Contaminants}
\label{sec:LY} 

An effect on the light yield in 100 Torr CF$_4$ was observed in our $^{55}$Fe data when acquired over many days.  This effect was due to contamination from out-gassing, which we discuss here.

$^{55}$Fe calibrations were required over 8 days for a different application\footnote{These were acquired for calibration of neutron exposures of the detector for directional studies.}, providing us data to study the effects of possible gas contaminants on the light yield.  Before the start of the 8 day data run, the detector vessel was sealed and pumped out to $\sim10^{-3}$ Torr overnight.  Subsequently, each day of data taking started by back-filling the detector vessel to 100 Torr with 99.999\% pure CF$_4$, powering up the detector and acquiring $\sim$70 minutes of $^{55}$Fe data.  This was followed by $\sim$14 hours of neutron data and a second $\sim$70 minute $^{55}$Fe exposure at the end of the day, after which the detector was powered down and the vessel pumped out for $\sim$3 hours.  This procedure was repeated for each day of data taking.  The GEMs were powered up to the same voltages each day, with the values given in Section~\ref{sec:gain} and Table~\ref{tab:config2} for the moderate, stable gas gain of $\sim1$$\times10^5$.  The $^{55}$Fe spectrum obtained from the charge signal was used to monitor the gas gain, which remained constant within 10\% over the 8 days.  The $^{55}$Fe spectrum obtained from the optical signal 
over this period, however, showed an increase in the light yield, which is highly sensitive to gas purity (e.g., \cite{margatopurity}).  With the detector kept sealed over the 8 days, out-gassing would be the main source of contamination with the expectation that its rate would decrease over time.

\begin{figure}[]
	\centering 
	\includegraphics[width=0.48\textwidth]{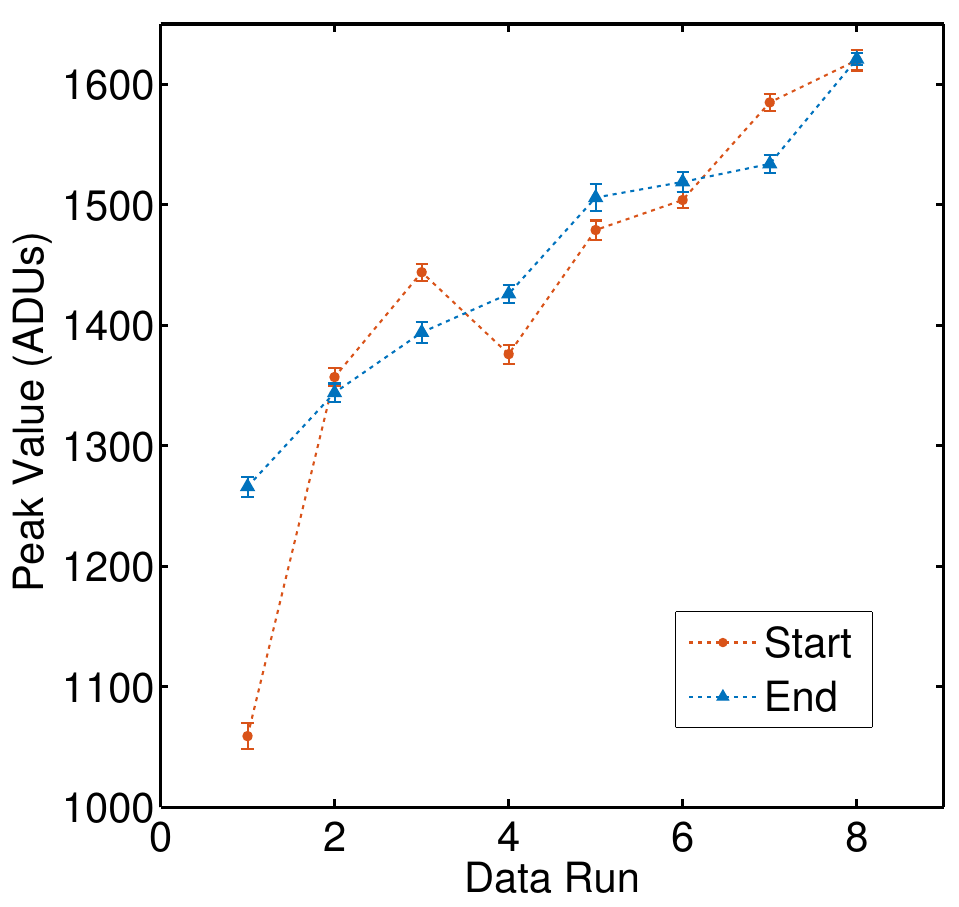}
	\caption{The fitted peak value of the $^{55}$Fe energy spectrum at the start and end of the day over eight days of data taking.  With the exception of the first data run, the start and end peak values are within 4\%.  The large change ($\sim 20\%$) seen between the start and end in the first data set is likely due to insufficient time for the GEMs to charge up and stabilize.  The raising peak value with run number is likely the result of a reduction in concentration of contaminants with the additional pump down in-between runs.  }
	\label{fig:peak_vs_days}
\end{figure}

The results of the data runs are plotted in Figure~\ref{fig:peak_vs_days}, which shows the fitted peak value of the optical $^{55}$Fe spectrum at the start and end of each of the 8 days of data acquisition.  The large difference in the start and end spectrum peak values for data run 1 is most likely the result of insufficient time given between GEM power up and data taking; we discovered that the GEMs required about 1-2 hours after being brought up to voltage before stable operation ensued.

With the exception of run 1, the start and end spectrum peak values are always within 4\% of each other, which suggests that any compositional change of the gas does not affect the light output significantly over half a day.  However, over 8 days the curves in Figure~\ref{fig:peak_vs_days}  show a gradual rise in the light output.  Between run 2 and run 8 the light output increased by 21\% in the end series, and by 19\% in the start series.  The light yield in the end series is monotonically increasing whereas that in the start series has more scatter. This is due to the variability in the time between GEM power up and data taking, as discussed above.  Nevertheless, the steady increase in light yield over the 8 days is consistent with improved gas purity. We attribute this to a decrease in the out-gassing rate, which can have a long half-life in sealed vessels that have not been baked out.

\section{Conclusion}
\label{sec:Conclusion}

We have shown that a GEM and THGEM based TPC detector can be operated in low pressure CF$_{4}$ (25-100 Torr) with effective gas gains exceeding $2 \times 10^5$.  With the high signal-to-noise and spatial resolution of this detector, it was possible to image and resolve individual $^{55}$Fe tracks  using an optical readout consisting of a fast lens and a low noise CCD camera.  $^{55}$Fe spectra were derived using the optical signal from these tracks, demonstrating good energy resolution with FWHM down to $\sim$$30\%$.  These results show that low pressure gas TPCs with optical readout could provide an interesting alternative to, and potential advantages over, traditional charge readout.  Specific areas that could benefit from this work are X-ray polarimetry and directional dark matter detection experiments, both requiring detailed track reconstruction of ionizing particles  down to the lowest energies.

\section*{ Acknowledgements}
\noindent This material is based upon work supported by the NSF under Grant Nos. 0548208, 1103420, and 1407773.


\end{document}